\documentclass{aa}
\usepackage{graphicx}
\include{epsf}
\def\lesssim{\mathrel{\hbox{\rlap{\hbox{\lower4pt\hbox{$\sim$}}}\hbox{$<$}}}}
\def\gtrsim{\mathrel{\hbox{\rlap{\hbox{\lower4pt\hbox{$\sim$}}}\hbox{$>$}}}}

\begin{document}


%
     \title{Two-dimensional Distributions and Column Densities of Gaseous
               Molecules in Protoplanetary Disks II.}

     \subtitle{Deuterated Species and UV Shielding by Ambient Clouds}

     \author{Y. Aikawa
            \inst{1}
            \and
            E. Herbst\inst{2}
            }

     \offprints{Yuri Aikawa}

     \institute{Department of Earth and Planetary Sciences, Kobe University,
                Kobe 657-8501, JAPAN\\
                email: aikawa@jet.planet.sci.kobe-u.ac.jp
           \and
                Departments of Physics and Astronomy, The Ohio State
                University,\\ Columbus, OH 43210, USA\\
                email: herbst@mps.ohio-state.edu
               }

     \date{Received; accepted}

\abstract{
We have investigated the two-dimensional ($R,Z$) distribution of
deuterated molecular species in circumstellar disks around young
stellar objects.  The abundance ratios between singly deuterated and
normal molecules (``D/H ratios'') in disks evolve in a similar way as
in molecular clouds.  Fractionation is caused by rapid exchange
reactions that are exothermic because of energy differences between
deuterated and normal species.  In the midplane region, where
molecules are heavily depleted onto grain surfaces, the D/H ratios of
gaseous molecules are higher than at larger heights.
The D/H ratios for the vertical column
densities of NH$_3$, H$_2$O, and HCO$^+$ are sensitive to the
temperature, and decrease significantly with decreasing radial
distance for $R\lesssim 300$ AU. The analogous D/H ratios for CH$_4$
and H$_2$CO, on the other hand, are not very sensitive to the
temperature in the range ($T=10-50$ K) we are concerned with, and do
not decrease with decreasing $R$ at $R\ge 50$ AU. The D/H column-density
ratios also depend on disk mass.  In a disk with a larger mass, the
ratios of deuterated species to normal species are higher, because of
heavier depletion of molecules onto grains.
In the second part of the paper, we report molecular column densities
for disks embedded in ambient cloud gas.  Our results suggest that CN
and HCO$^+$ can be tracers of gaseous disks, especially if the central
object is a strong X-ray source .
Our results also suggest that the radial distributions of CN,
C$_2$H, HCN, and H$_2$CO may vary among disks depending on the X-ray
luminosity of the central star.
}

\titlerunning{Molecular evolution in protoplanetary disks. II}
     \maketitle

\section{INTRODUCTION}

Protoplanetary disks are formed around young stellar objects as the
stars are created, and are birthplaces of planetary systems.
Molecular evolution in protoplanetary disks is important because it
can reveal the chemical connection between planetary and interstellar
matter. Since chemical abundances are determined by physical parameters such
as temperature and density, studies of molecular evolution in protoplanetary
disks can also be of help in determining the structure and physical evolution
of the disks from the molecular line observations.

In recent years it has become possible to detect molecular lines in
protoplanetary disks by radio astronomy.  Aperture-synthesis images
have directly revealed gaseous CO disks, which are in Keplerian
rotation (Kawabe et al.  1993; Koerner et al.  1993; Dutrey et al.
1994; Koerner \& Sargent 1995; Saito et al.  1995; Guilloteau \&
Dutrey 1998).  In addition to the CO studies, Dutrey et al.  (1997)
surveyed other molecular lines in the disks around DM Tau and GG Tau
using the IRAM 30 m telescope.  These stars, with ages $1\; 10^6$ yr
and $3\; 10^5$ yr respectively (Beckwith et al.  1990; Handa et al.
1995), have large gaseous disks with radii $\sim 800$ AU.  The molecular
line spectra detected towards the two
stars are similar despite the fact that the disk around GG tau is
circumbinary.  The averaged fractional abundances of gaseous CO, CN, CS,
HCN, HNC, H$_2$CO, C$_2$H, and HCO$^+$ were reported by Dutrey et al.
(1997), who found that the abundances of heavy-element-containing molecules
relative to hydrogen are lower than those in molecular clouds by
factors of 10-100, and relative abundances among heavy molecules are
also different.  Specifically, the abundance ratio of CN to HCN is
much higher in disks than in molecular clouds.

In order to reveal the chemical and physical causes for these
differences, Aikawa \& Herbst (1999a) (hereinafter Paper I)
investigated the two-dimensional distribution of molecules in a disk
by calculating molecular concentrations from a network of chemical
reactions.  They found molecular abundances to vary with height $Z$
from the midplane.  At larger radii ($R\gtrsim 100$ AU), which
contribute most to the molecular emission lines because of the large
area, the temperature is so low ($T\sim 10$ K), that most molecules,
except for H$_2$ and He, are partially adsorbed onto grains to form
ice mantles.  In the midplane region ($Z \approx 0$) molecular
depletion is the most effective because disks are in hydrostatic
equilibrium in the vertical direction, and the density is the highest
in the midplane region.

In regions above and below the midplane, significant amounts of
molecules remain in the gas phase because of lower densities, and
because of non-thermal desorption, which is caused by cosmic rays
and/or radiation (ex. X-rays) from the interstellar field and
the central star.  Aikawa \&
Herbst (1999a) suggested that the observed molecular lines come mostly
from this region.  There is a height distinction between stable
molecules and radicals, however.  In the surface (uppermost) region of
a disk, radicals such as CN are very abundant because of
photodissociation via UV radiation.  The molecular column densities
obtained in Paper I by integrating over height are in reasonable
agreement with observation; molecular depletion in the midplane
explains the low average abundance of heavy-element-containing species
relative to hydrogen, and the high abundance ratio of CN to HCN is
caused by photochemistry in the surface region.  It is clear that
consideration of the 2-D distribution of molecules is essential in
order to interpret the observed molecular line intensities.

In this paper we present two extensions of the work in Paper I.
Firstly, we include deuterium chemistry in the 2-D disk model.  Since
the line survey by Dutrey et al.  (1997), further searches for
molecular lines in protoplanetary disks have been conducted by several
groups of observers.  One of the most interesting results is the
detection of deuterated species in the disk of LkCa15 (Qi et al. 1999;
Qi 2000).
The average DCN/HCN ratio is estimated to be about 0.01, which is
much higher than the D/H elemental abundance ratio of $1.5\;
10^{-5}$.  Aikawa \& Herbst (1999b) showed that a high D/H ratio in a
disk is a natural outcome of the incorporation of interstellar
material to the protoplanetary disk and a low temperature disk
chemistry, which is similar to cloud chemistry.  They were concerned,
however, only with icy material in the midplane region, and did not
predict column densities for deuterated gaseous molecules.  A
two-dimensional disk model with deuterium chemistry and comparison
with observation should be helpful in constraining the physical properties
of the disk, such as temperature.  Such a theoretical model should also
be useful in guiding searches for other deuterated molecules.
The results presented here fulfill these expectations, at least
partially.

Secondly, we consider disks that are embedded in molecular clouds or
circumstellar envelopes.  Up to now, molecular line observations of
disks have been successful only in cases where a disk is removed from
molecular clouds or where the system velocity is quite different from
the cloud velocity.  Most disks, which presumably are embedded in
cloud gas or circumstellar envelopes, are hard to observe via
molecular lines because of contamination with ambient gas.  If we were
able to predict which molecules could selectively trace disk gas, the
species could be used to search for embedded gaseous disks, and so
permit more statistically reliable arguments on disk properties, such
as their size and time scales of gaseous dissipation.

The rest of the paper is organized as follows.  In \S 2 we describe
our model of protoplanetary disks and the chemical reaction network
we utilize.  Numerical results on the distribution of molecular
abundances, the column density ratio of deuterated and normal
species, and the dependence of D/H ratios on selected physical
properties of disks are discussed in \S 3.  In \S 4, we report our
investigation of molecular abundances in disks that are embedded in
ambient gas, and discuss which, if any, molecular lines can be used
as disk tracers.  We summarize our results in \S 5.

\section{Model}
\subsection{Disk Model}
As in Paper I, we have adopted simple static disk models
-- the Kyoto model (Hayashi 1981) extrapolated to 700 AU, and a model
in which the mass and density are lowered by an order of magnitude.
The lower mass model is similar to that adopted by Dutrey et al. (1997),
with which they estimate the averaged molecular abundances in DM Tau.
The total disk mass is about $6\; 10^{-2} M_{\odot}$ for the extended
Kyoto model, and $6\; 10^{-3} M_{\odot}$ for the lower mass model.
Although much work has been undertaken to formulate the structure and
physical evolution of protoplanetary disks (Cameron 1973; Hayashi
1981; Adams \& Lin 1993, and references therein), there is still no
obvious standard model.  The main goal of
this paper is to comprehend the essential characteristics of deuterium
chemistry in a protoplanetary disk using simple disk models.

The details of the Kyoto model are described in Paper I.
The column density of hydrogen nuclei
$\Sigma_{\rm H}$  (cm$^{-2}$) and the temperature $T$(K) as functions
of $R$ are given by the equations
\begin{eqnarray}
\Sigma_{\rm H}(R)&=&7.2\; 10^{23} \left(\frac{R}{100 \rm AU}\right)^{-3/2}
\label{eq:sigma}\\
T(R)&=&28\left(\frac{R}{100 {\rm AU}} \right)^{-1/2}\left(\frac{L_{\ast}}
{L_{\odot}}\right)^{1/4}  .
\label{eq:temp}
\end{eqnarray}
The luminosity of the central star $L_{\ast}$ is assumed to be
1 $ L_{\odot}$ in this paper.
It is assumed for simplicity that the disk is isothermal
at each radius.

The gas is in hydrostatic equilibrium in the vertical direction.
  From the mass distribution given by equation (\ref{eq:sigma}) and
the temperature distribution given by equation (\ref{eq:temp}), the density
distribution by number of hydrogen nuclei (cm$^{-3}$) can be shown to be
\begin{eqnarray}
n_{\rm H}(R,Z)&=&1.9\; 10^9 \left(\frac{R}{100 {\rm AU}}\right)^{-11/4}
{\rm exp}\left(-\frac{GM_{\ast}\mu m_{\rm H}}{RkT}\right) \nonumber \\
&\times&{\rm exp}\left[\frac{GM_{\ast}\mu m_{\rm H}}{kT(R^2+Z^2)^{1/2}}\right]
\label{eq:density}
\end{eqnarray}
where $G$ is the gravitational constant, $\mu$ is the mean molecular
weight of gas, $m_{\rm H}$ is the mass of a hydrogen atom, and $k$ is the
Boltzmann constant. Since the disk mostly consists of H$_2$ and He,
the mean molecular weight $\mu$ is 2.37, with the
elemental abundances of Anders \& Grevesse (1989). The mass of the
central star $M_{\ast}$ is assumed to be $1M_{\odot}$.

\subsection{Reaction Network}
The basic equations for molecular evolution are given by
\begin{equation}
\frac{dn(i)}{dt}=\sum_{j} \alpha_{ij}n(j)
+\sum_{j,k} \beta_{ijk}n(j)n(k),
\label{eq:reaction}
\end{equation}
where $n(i)$ is the number density of species $i$,
and the $\alpha_{ij}$ and $\beta_{ijk}$ are rate coefficients.
We use the ``new standard model'' network of
chemical reactions for the gas-phase chemistry (Terzieva \& Herbst
1998; Osamura et al.  1999).
The ionization rate by cosmic rays is assumed to be the ``standard'' value
for molecular clouds, $\zeta =1.3\; 10^{-17}$ s$^{-1}$ (e.g. Millar et al.
1997), because
the attenuation length of cosmic ray ionization is much
larger than the column density for $R\ge 50$ AU. The ionization rate 
is uncertain by a factor of a few at least (e.g. van der Tak \& van 
Dishoeck 2000).  Calculations with $\zeta =2.6\; 10^{-17}$ s$^{-1}$ 
yield molecular column densities that differ by less than a factor of
3 from those with our standard $\zeta$.

 We have extended
the network to include mono-deuterated analogues of hydrogen-bearing
species (Millar et al. 1989; Aikawa \& Herbst 1999b). For normal
exothermic reactions and dissociative recombination reactions,
we have assumed that the total rate coefficient is unchanged
for deuterated analogues, and have also assumed statistical branching
ratios. There are some exceptions to the statistical rules; for
example,
the dissociative recombination of HCND$^+$ does not produce
DCN but HCN. Similar rules are set for the hydrogenation of HCN and DCN,
i.e., HCN + H$_2$DO$^+$ produces HCND$^+$ or HCNH$^+$, but not DCNH$^+$.
Another important exception is
\begin{eqnarray}
{\rm H_2CN} + {\rm D} & \to & {\rm HCN} + {\rm HD} ~ ~ (a) \nonumber \\
              & \to & {\rm DCN} + {\rm H_{2}}  ~ ~ ~ ~ (b).  \nonumber
\end{eqnarray}
The lower branch ($b$) is one of the main formation paths for DCN. If we apply
the statistical rule, the branching ratio $a/b$ is 2. However,
the experiment by Nesbitt et al. (1990) shows the ratio
$a/b=5\pm 3$. In this paper we assume the ratio $a/b$ to be 5.

We have included those deuterium exchange reactions for molecular
ions and HD that are known to proceed in the laboratory or have been
studied in detail theoretically (Millar et al. 1989); such reactions
drive the fractionation yet are known to occur for only a few ions
since activation energies are common (Henchman et al. 1988).

Although our model does not contain surface chemistry, as does the
latest model of Willacy \& Langer (2000), we do include the surface
formation of
H$_2$ molecules, the surface recombination of ions and electrons, and
formation and desorption of ice mantles. As in Paper I, we adopt an
artificially low sticking probability $S=0.03$ for adsorbing
species on the grain surface, in order to mimic the effect of
non-thermal desorption. The sticking coefficient was originally
chosen to fit the observed spectrum of CO emission lines in GG Tau.
For thermal  desorption
from ice mantles, we adopt the same rate coefficients as in Aikawa et
al.  (1997).  The total numbers of
species and reactions included in our network are 773 and 10539,
respectively.

The elemental abundances used here are the so-called ``low-metal'' values
(e.g.  Lee et al.  1998; Aikawa et al.  1999).  The initial molecular
abundances are determined by following
molecular evolution in a precursor molecular cloud core with physical
conditions $n_{\rm H}=2\; 10^4$ cm$^{-3}$ and $T=10$ K up to
$3\; 10^5$ yr, at which time observed abundances in molecular
clouds are reasonably reproduced (Terzieva \& Herbst 1998).

\subsection{X-rays}
Results from the X-ray observation satellites, ROSAT and ASCA, show
that T Tauri stars are strong X-ray emitters, with X-ray luminosities
   in the range $10^{29}-10^{31}$ erg s$^{-1}$ (Montmerle et al. 1993;
Glassgold et al. 1997). We include chemical processes caused by X-rays
following Maloney et al. (1996) and Glassgold et al. (1997).
In Paper I, we showed that X-rays affect the disk chemistry via ionization
and induced UV radiation.
Direct and secondary ionization of heavy elements by X-rays, which
were not taken into account in Paper I, are included in this paper.
We find that inclusion of these new processes does not much alter the results
of Paper I. For completeness, however, we discuss their inclusion. We adopt
an X-ray luminosity of $10^{31}$ erg s$^{-1}$,
which is almost the upper limit of the observed luminosity, in order to
examine the upper limit of the X-ray effects.

\subsubsection{Secondary ionization}
X-rays produce electrons with energies of several hundred eV
by ionizing atoms and molecules. These electrons cause secondary
ionization of about 30 molecules and atoms per
keV of primary electron energy, most of which are hydrogen molecules. 
Secondary ionization
is more effective than direct
ionization by X-rays in a hydrogen-dominated gas, so that the
overall ionization rate per unit time
is given approximately by
\begin{equation}
\zeta_{\rm x}=N_{\rm sec}\int \sigma(E)F(E)dE,
\end{equation}
where $N_{\rm sec}$ is the number of secondary ionizations per unit 
primary photoelectron energy, $\sigma$ is the cross section for direct X-ray
ionization for all elements weighted by their solar abundances,
and the X-ray photon flux $F(E)$, in units of per unit area per 
unit time, is given by
\begin{equation}
F(E)=F_{\rm o}(E)e^{-\tau(E)}
\label{eq:xflux}
\end{equation}
where $\tau$ is the optical depth along a path from the central star. 
The actual value of $\zeta_{\rm x}$ at a radius of $R=700$ AU in the Kyoto model
is shown as a function of height from the midplane in Figure 2 of Paper I.
The ionization rate via X-rays is higher than that via cosmic rays in 
regions roughly above the scale height of the disk.

Although the photoelectrons ionize mostly H$_2$ and H,
the secondary ionization of heavy elements may be
important for their chemistry.  Secondary ionization rate coefficients
for heavy elements are estimated by
\begin{equation}
\zeta_{\rm x}^{\rm m}=\zeta_{\rm x}
     \frac{\sigma_{\rm ei, m}(E)}{\sigma_{\rm ei, H}(E)}
\end{equation}
where $\sigma_{\rm ei, m}(E)$ is the electron-impact ionization cross section
of element $m$ at energy $E$. We used the cross sections given by
Lennon et al. (1988), and averaged the ratio  $\sigma_{\rm ei, m}(E)/
\sigma_{\rm ei, H}(E)$ over the energy range $0.1-1$ keV, because
the energy of the primary electron is a few hundred eV. Typically, the ratio
is $\sim 3-7$ for elements such as C (4.29) and Mg (7.09).

\subsubsection{Direct ionization}

The direct X-ray ionization
rates for C, N, O, Si, S, Fe, Na, Mg, Cl and P are calculated at each point
of the disk using the X-ray flux obtained from  equation (\ref{eq:xflux})
and the ionization cross sections given by Verner et al. (1993).
We assume primary ionization
of heavy elements in a molecule leads to a doubly ionized species because of
the Auger effect; the doubly ionized species then dissociate into
two singly charged ions.
We consider this destructive reaction only for simple diatomic molecules,
because we do not know the products when polyatomic molecules are
dissociated by X-rays. This simplification does not affect our
results, because X-ray induced photolysis is much more efficient than direct
ionization in terms of molecular destruction.

\subsubsection{UV photolysis induced by X-rays}
Energetic photoelectrons produced by X-rays collide with hydrogen atoms
and molecules, and generate UV photons, just as cosmic rays do.
Although this process was considered in Paper I, it was not discussed.
In H$_2$-dominated regions, the photoreaction rate coefficient
for a species $m$ is given by
\begin{equation}
R_{\rm m}=2.6 \psi p_{\rm m} x_{\rm H2} \zeta_{\rm x} (1-\omega)^{-1}
~ ~ {\rm s}^{-1}
\label{eq:inducedUV}
\end{equation}
where $\psi = 1.4$ is the ratio of number of Lyman-Werner photons
produced per H$_2$ ionization (Maloney et al. 1996).
Since hydrogen is mostly in the molecular form in the region we are
interested in, we adopt the constant value $x_{\rm H2}=0.5$ in this formula.
We can ignore the contribution from hydrogen atoms because
atomic hydrogen only dominates in the surface regions, which
do not contain many molecules.
The coefficient $p_{\rm m}$ is given by Gredel et al. (1989), and $\omega$
is the grain albedo, which is assumed to be 0.5. For molecules which are
not listed in Gredel et al. (1989), we estimate the value of $p_{\rm m}$
from similar molecules. Following Gredel et al. (1987), we adopt
$p_{\rm m}=10$ for CO. The deuterated hydrogen molecule HD is also
subject to photodissociation by induced UV. Because HD is dissociated
by lines, just like CO, we assume $p_{\rm m}=10$ for HD.

In fact, equation (\ref{eq:inducedUV}) assumes that induced photons are
absorbed by molecules locally; in other words, a spatial gradient of
$\zeta_{\rm X}$ is not considered. Since $\zeta_{\rm X}$ is higher at
larger heights in our model, and
since induced photons can be emitted in the vertical direction of the disk,
the induced photons could penetrate deeper into the disk than we
estimate, especially in the outer radius ($R\gtrsim 500$ AU) in our lower
mass disk, in which the total column density is low ($A_{\rm v}\lesssim
1$ mag). But a quantitative calculation of this effect requires a 2-D
radiation transfer calculation, which is beyond the scope of this paper.

\subsection{UV Radiation}

A protoplanetary disk is irradiated by UV radiation from the external
interstellar field and from radiation due to the central star.  The UV
flux from the central star varies temporally, and at $R=100$ AU the
unattenuated UV flux can reach a value $10^4$ times higher than the
interstellar flux (Herbig \& Goodrich 1986; Imhoff \& Appenzeller
1987; Montmerle et al.  1993).  As in Paper I, we utilize this maximum
value.  The radiation fields from the two sources strike the disk from
different directions, and thus suffer different degrees of
attenuation.  We obtain the attenuation of interstellar UV in terms of
the visual extinction $A_{\rm v}$, by calculating the vertical column
density from the disk surface to the height we are interested in,
using the relation $A_{\rm v}= N_{\rm H}/[1.8 \; 10^{21}$ cm$^{-2}$
mag$^{-1}$].  The attenuation of the stellar UV, $A_{\rm v}^{\rm star}$,
is obtained by calculating the column density from the central star.
Values of $A_{\rm v}$ and $A_{\rm v}^{\rm star}$ are given in
Table 1 of Paper I, as a function of height at $R=700$ AU for the
Kyoto model.  These values of extinction are then put into the
photo-rate equations of our network.

In \S 4, where we report calculated values for molecular
distributions in embedded disks, we consider an additional attenuation
of $1-2$ mag via ambient gas, while the attenuation of the stellar
radiation is not modified.

Although the UV radiation is mainly attenuated by dust, self- and
mutual-shielding must be considered for H$_2$ and CO (van Dishoeck \& Black
1988; Lee et al.  1996), so that we must solve for
the molecular abundances and the UV attenuation self-consistently. As in
Paper I, we solve a one-dimensional slab model at each radius of the disk
and utilize modified shielding factors from Lee et al. (1996).

\section{Results}
\subsection{Vertical Distribution}

\begin{figure*}
\resizebox{12cm}{!}{\includegraphics{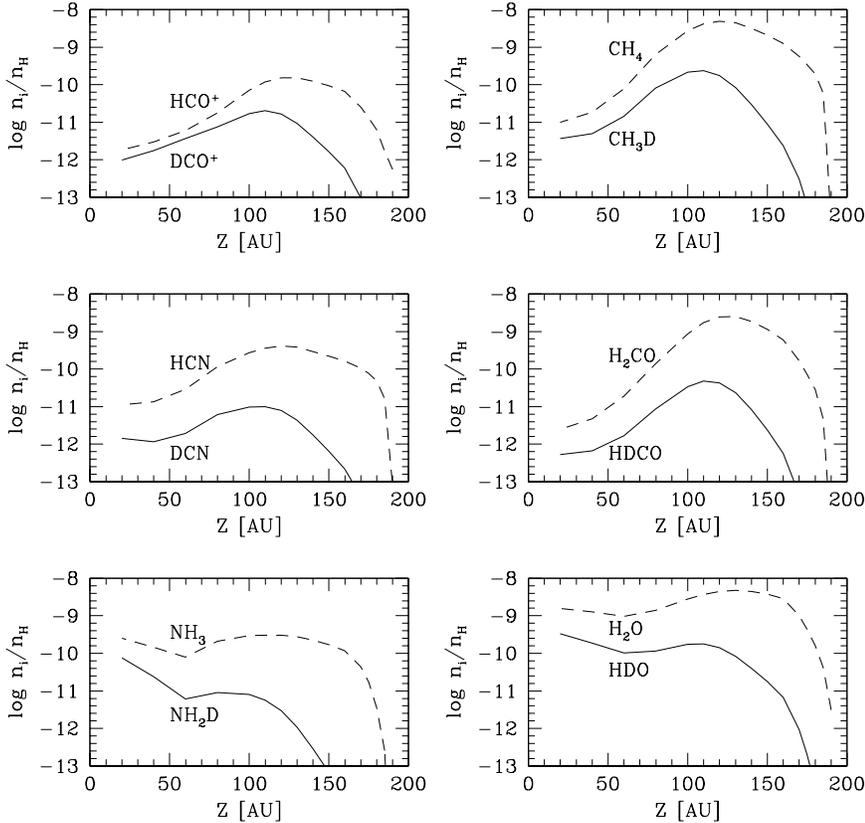}}
\caption{Vertical distribution of normal (dashed lines) and deuterated
(solid lines) molecules at $R=500$ AU. The disk age is assumed to
be $t=9.5\; 10^5$ yr and the disk mass is less massive than the
Kyoto model by an order of magnitude.}
\label{fig:vertical_rel}
\end{figure*}

Fig.  \ref{fig:vertical_rel} shows the vertical distribution of
various molecules (dashed lines) and their singly deuterated
counterparts (solid lines) at $R=500$ AU and at $t=9.5\; 10^5$ yr,
which is a typical age of a T Tauri star.  The disk is assumed to be
less massive than the Kyoto model by an order of magnitude.  The
effects of X-rays from the central star are included.  Throughout this
paper, we will also present the results of models in which the mass is
that of the Kyoto model, in which X-rays are turned off, and in which
the time is changed.  The low mass model is emphasized because in
Paper I, we showed that it is consistent with the observational
result of DM Tau by Dutrey et al. (1997).

As can be seen in Fig. \ref{fig:vertical_rel}, most molecules
have a peak abundance at some intermediate height; in the surface
region of the disk the molecules are dissociated by UV photons both
from the interstellar field and the central star, while close to the
midplane most molecules are adsorbed onto grains.  In the midplane
region, however, some gas-phase species show a ``late-time peak'' at
$t\sim 10^5-10^6$ yr (Ruffle et al.  1997; Paper I) which can occur in
gas-grain models with low sticking efficiencies or with non-thermal
desorption. The high abundance of NH$_3$ in the midplane is caused by
this peak.

A comparison of the concentrations of molecules and their deuterated
isotopomers in Fig. \ref{fig:vertical_rel} shows that molecular D/H ratios
are much higher
than the elemental abundance ratio D/H of $1.5\; 10^{-5}$. This
is not surprising because
deuterium fractionation proceeds in a similar way as in molecular
clouds.
Owing to the energy differences between deuterated species and normal species,
and to some rapid exchange reactions, species such as H$_3^+$ and CH$_3^+$
have a high D/H ratio, and the high ratio propagates to other species
through ion-molecule reactions (Millar et al. 1989; Aikawa
\& Herbst 1999b).
Another route to fractionation lies through the dissociative
recombination of molecular ions with high D/H ratios, which leads
to a high atomic D/H ratio.
Neutral-neutral reactions involving H and D then propagate these high
atomic D/H
ratios.  For example, a major production route for DCN is the
reaction between D and H$_2$CN.

Fig.  \ref{fig:vertical_rel} also shows that molecular D/H ratios are
higher at smaller heights, where molecules are more heavily depleted
from the gas phase onto grain surfaces.  Molecular depletion enhances
the D/H ratio of the remaining gaseous species.  For example,
H$_2$D$^+$ is formed by the reaction H$_3^+$ + HD and, in many
situations, is destroyed mainly by the
reaction H$_2$D$^+$ + CO. Hence the ratio H$_2$D$^+$/H$_3^+$ is
proportional to $n$(HD)/$n$(CO), which increases as the gaseous CO
abundance decreases (Brown \& Millar 1989).  At sufficiently small
CO densities, the dominant destruction route for H$_2$D$^+$ becomes
reaction with electrons or back reaction with H$_{2}$.

\subsection{Radial Distribution of D/H ratios}

\begin{figure*}
\resizebox{12cm}{!}{\includegraphics{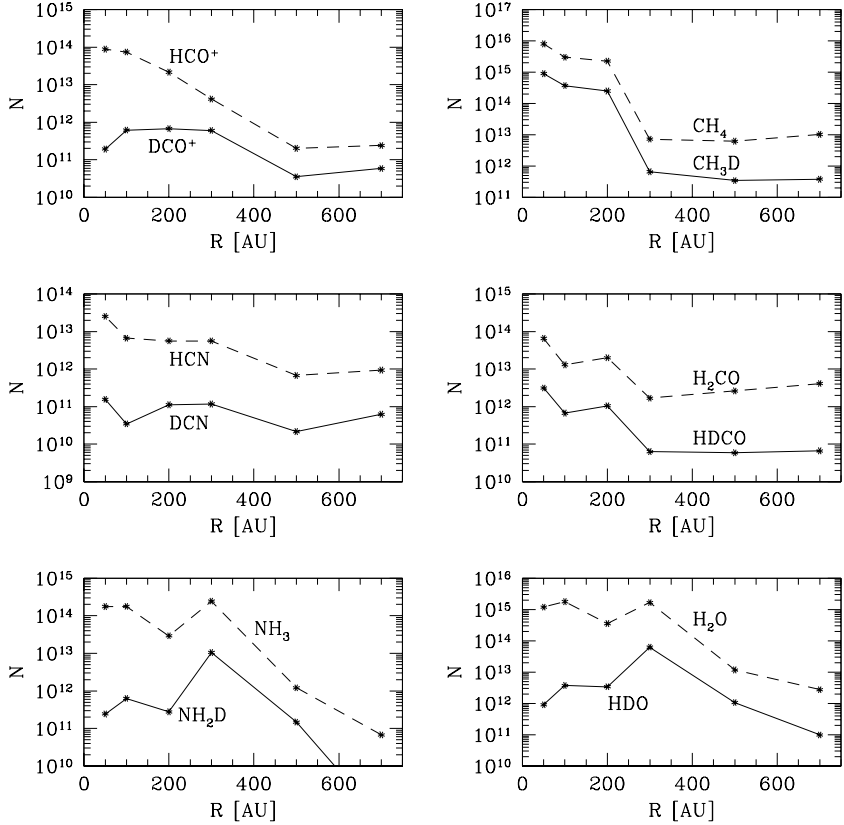}}
\caption{The column densities of deuterated and normal species
as a function of radius. The disk mass is assumed to be less than the
Kyoto model by an order of magnitude. The disk age is
$t=9.5\;10^5$ yr.}
\label{fig:column_01_16X}
\end{figure*}

\begin{figure*}
\resizebox{12cm}{!}{\includegraphics{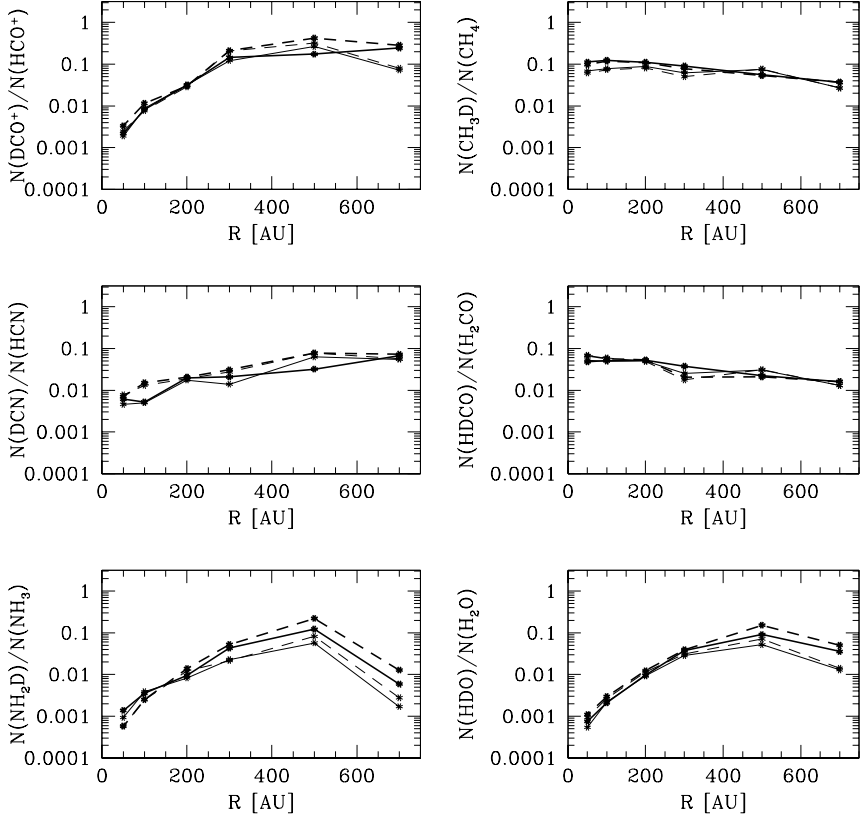}}
\caption{The column density ratios of deuterated to normal species as a
function of radius. The disk mass is assumed to be less than the Kyoto model
by an order of magnitude. The thick lines show the ratios at $t=9.5\; 10^5$
yr, the thin lines at $t=3\; 10^5$ yr. The solid lines show models
with X-rays and the dashed lines without X-rays.}
\label{fig:column_DH_01}
\end{figure*}

\begin{figure*}
\resizebox{12cm}{!}{\includegraphics{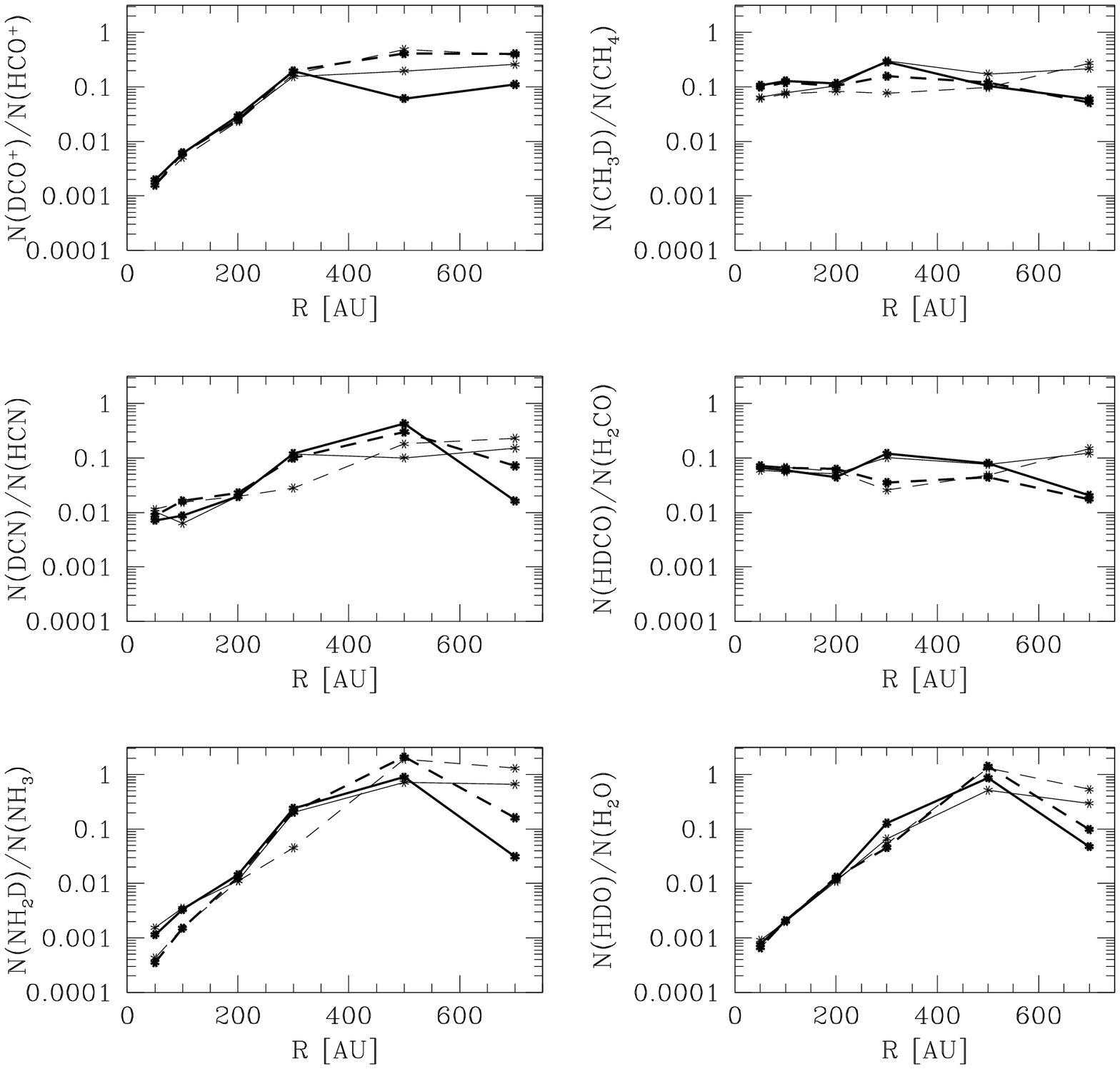}}
\caption{The column density ratios of deuterated to normal species
as a function of radius, but for the Kyoto model. The thick lines show the
ratios at $t=9.5\; 10^5$
yr, the thin lines at $t=3\; 10^5$ yr. The solid lines show models
with X-rays and the dashed lines without X-rays.}
\label{fig:column_DH_1}
\end{figure*}

By integrating the vertical distributions of molecular abundances,
we obtain molecular column densities. The column
densities of deuterated and normal species are shown in Fig.
\ref{fig:column_01_16X} for assorted molecules as a function of disk radius.
The time is the same as in Fig. \ref{fig:vertical_rel}:
$t=9.5\; 10^5$ yr.
Despite the addition of direct and
secondary ionization of heavy elements via X-rays in
this paper, the calculated column densities of normal species
are almost the same as in Paper I. One notable modification concerns the NH$_3$
column density in the outermost region ($R \gtrsim 500$ AU) of the disk.
In Paper I, we assumed that the H$_3^+$ + N reaction produces NH$^+$;
however, subsequent information shows that the reaction does not
proceed efficiently (Scott et al. 1998). In the outer region, where
N atoms are relatively abundant
because of the low density, this reaction was a key component in the
synthesis  of NH$_3$. At $R=700$ AU in the lower mass model with
X-rays and $t=9.5\;
10^5$ yr, the column density of NH$_3$ was $1.2\; 10^{12}$ cm$^{-2}$
in Paper I, but it is only $6.7\; 10^{10}$ cm$^{-2}$ in the current model.
At smaller radii ($R\lesssim 300$ AU) the modification is less significant.

We obtained averaged D/H ratios for molecules at each radius by dividing
the column density of the deuterated species by that of the normal species.
Fig. \ref{fig:column_DH_01} shows D/H ratios for the lower mass disk model, and
Fig. \ref{fig:column_DH_1} for the Kyoto model. The thick lines show
the ratio at $t=9.5\;10^5$ yr, and the thin lines at a shorter
time of $3\; 10^5$ yr.
The solid lines are for models with X-rays and the dashed lines for
models without X-rays.


The radial dependence of the column density ratios shown in Fig.
\ref{fig:column_DH_01} and \ref{fig:column_DH_1} can be understood
from the major mechanism of deuteration.  For example, the D/H ratios
of NH$_3$, H$_2$O, and HCO$^+$ decrease at $R\lesssim 300$ AU. These
three species are deuterated through H$_{2}$D$^+$.  Since the
exothermicity for the reaction H$_3^+$ + HD $\to$ H$_2$D$^+$ + H$_2$
is relatively low (230 K), the back reaction becomes important as the
temperature rises near the star and lowers the D/H ratio in the inner
regions.  On the other hand, the D/H ratios in CH$_4$ and H$_2$CO do
not decrease inwards, because those species are deuterated through
CH$_{2}$D$^+$, for which the exothermicity of the deuterium exchange
reaction is much higher (370 K) than that for H$_2$D$^+$.  Finally, the
DCN/HCN column density ratio slightly decreases inwards because  D
atoms are less abundant in inner regions.

X-rays affect the column density ratios. For $R\gtrsim 300$ AU,
the D/H ratios of HCO$^+$, NH$_3$ and H$_2$O are smaller in the case
with X-rays than otherwise, because the enhancement of the H$_2$D$^+$
to H$_3^+$ ratio is limited by the rate at which H$_2$D$^+$ is destroyed
by recombination with electrons, which are
more abundant in the case with X-rays (Gu$\acute{\rm e}$lin et al.
1982; Caselli et al. 1998). In the inner regions, on the other hand,
H$_2$D$^+$ is always destroyed more efficiently by the reaction with
CO or H$_2$, and thus the ratios are less dependent on the electron abundance.

The D/H ratios also depend on the total mass of the disk. Comparing Fig.
\ref{fig:column_DH_01} with Fig. \ref{fig:column_DH_1}, we can see that,
   except for DCO$^+$/HCO$^+$, the ratios are higher in the Kyoto model,
especially in the outer region. This mass dependence is
caused by molecular depletion onto grains, which is more efficient in
disks with higher mass.
This dependence does not appear for HCO$^+$, because its abundance peaks at
a larger height than those of neutral species, at which D/H ratios are less
affected by depletion.

So far, detection of deuterated species has been reported only
in the disk around LkCa15, where Qi (2000) observed DCN and HDO
using the OVRO interferometer (see also Qi et al. 1999). He found
DCN and HCN to be distributed within a radius of $\sim$ 1''
(140 AU at the distance of 140 pc) and $\sim$ 3''-4'' from the central
star, respectively. The differing sizes of the distributions
reflect the fact that the only DCN line detected is the $J=3-2$
transition. The column density of DCN was estimated to be $1\;
10^{13}$ cm$^{-2}$ from the integrated intensity of DCN($J=3-2$),
while the column density of HCN was estimated to be $\sim
10^{15}$ cm$^{-2}$ from the intensities of the H$^{13}$CN $J=3-2$ and
$J=1-0$ lines.  Our model result for the DCN/HCN ratio within
$R\lesssim 200$ AU is consistent with the observation, independent of
the disk mass.  However, the absolute column densities of HCN and DCN
in the model are significantly smaller than observed.  In the region
of radius $\lesssim 200$ AU, the column density of DCN is $\sim
10^{11}$ cm$^{-2}$ in our low mass disk model (Fig.
\ref{fig:column_01_16X}) and is $\lesssim 10^{12}$ cm$^{-2}$ in the
Kyoto model (see Fig.  \ref{fig:column_DH_1} and Paper I).  
Moreover, the column densities of various molecules detected in LkCa15
are significantly higher than those in DM Tau (Qi 2000). One possible
explanation for this difference would be the disk mass. The mass of
the disk around LkCa15 is estimated to be 0.2 $M_{\odot}$ from the
dust continuum, which traces the region of radius $\sim 100$ AU, while
the mass of Kyoto model within 100 AU is 0.024 $M_{\odot}$.
In addition to modifying the
Kyoto model to include higher masses, the major possibilities for
improving our absolute column densities are (a) to lower the
artificial sticking
probability used in our current model (S=0.03) so that more
material remains in the gas phase at a given time, or (b) to include
specific non-thermal desorption mechanisms (Willacy \& Langer 2000),
and to consider the variability of their efficiency among disks.
Finally, a K5 star HD284589, located close to LkCa15, may affect the molecular
abundances in the disk through heating and/or UV radiation.
These are prospects for future work. Observations of deuterated
species in DM Tau would be desirable to compare with our
current calculated values.

Like DCN, the estimated column density of HDO in LkCa15 -- $(2-7)\; 10^{14}$
cm$^{-2}$ (Qi 2000) -- is higher than the value ($\lesssim 1\;
10^{14}$ cm$^{-2}$) obtained in our models.  Since H$_2$O cannot be
observed from the ground, the HDO/H$_2$O ratio is not determined.
It is interesting that the intensity peak of HDO is offset from the
central star, which is consistent with our model.

\section{Disks Embedded in Ambient Gas}
\begin{figure*}
\resizebox{12cm}{!}{\includegraphics{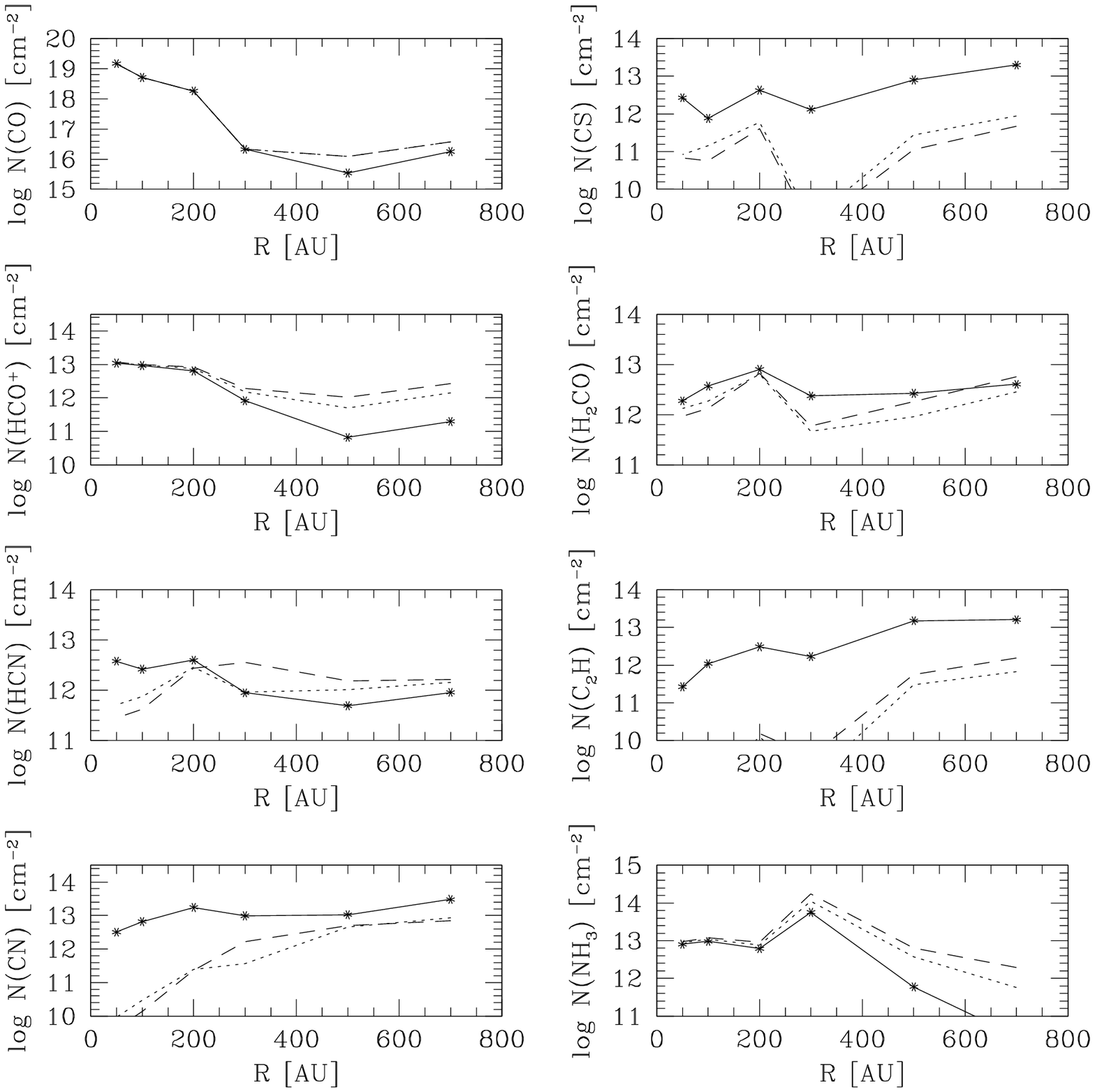}}
\caption{Column densities of assorted molecules as a function of radius.
The disk mass is assumed to be less than the Kyoto model
by an order of magnitude, and the disk age is $t=9.5\; 10^5$ yr.
Ionization, dissociation, and induced photolysis by X-rays are not
considered.
The solid lines show results for disks directly exposed to the interstellar UV
field, while the dashed and dotted lines show results for disks  embedded
in ambient gas of $A_{\rm v}=1$ and 2, respectively.}
\label{fig:atten_noX}
\end{figure*}

\begin{figure*}
\resizebox{12cm}{!}{\includegraphics{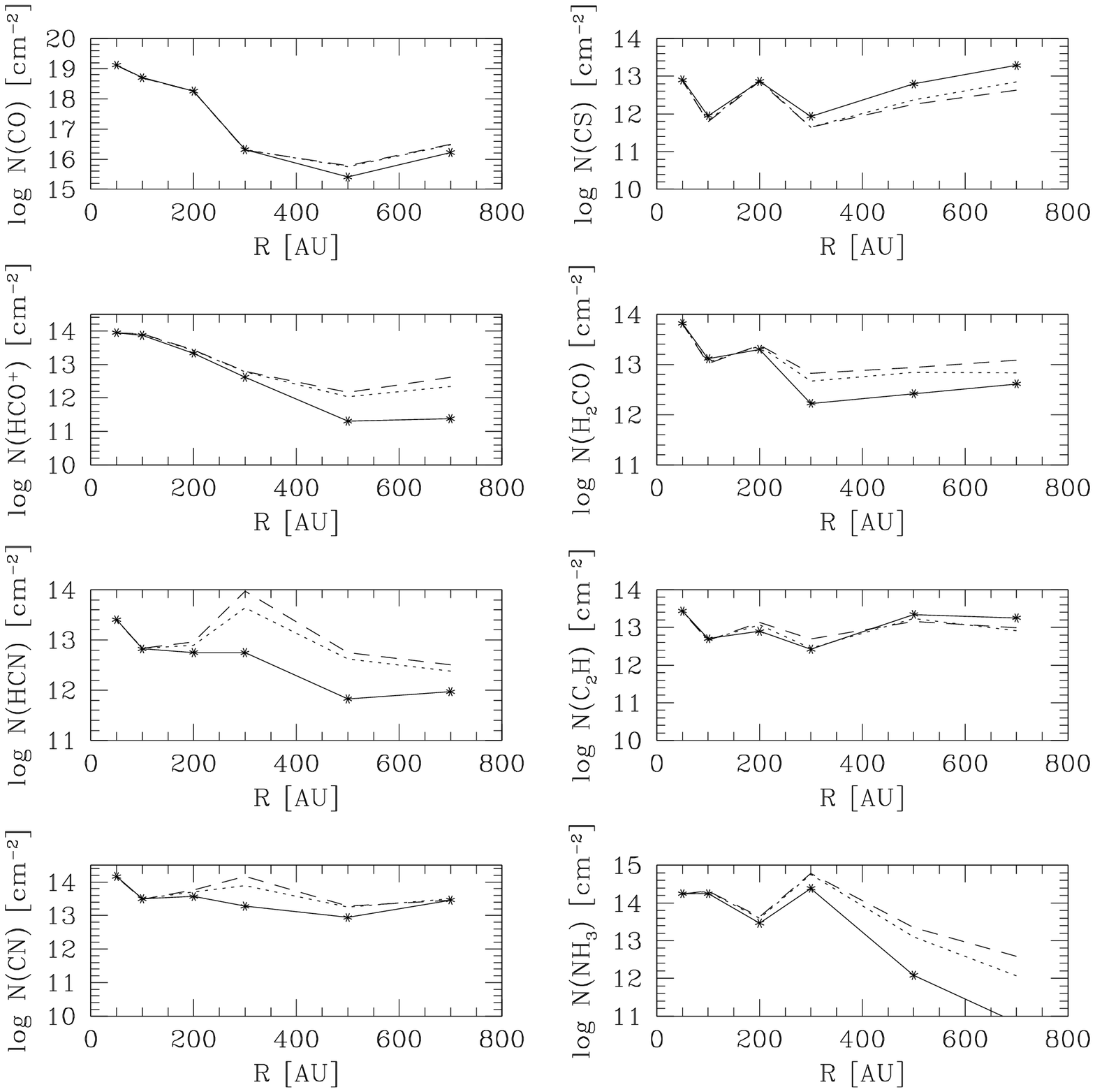}}
\caption{Same as in Fig. \ref{fig:atten_noX}, but with X-rays from the
central star included.}
\label{fig:atten_X}
\end{figure*}

Statistical observations of molecular emission lines in disks are
important in order to determine their size, time scale of gas
dissipation, and chemical abundances. Several gaseous disks have 
been
found via surveys of CO $J=2-1$ and $J=1-0$ lines (Kawabe et al.
1993; Koerner et al. 1993; Dutrey et al. 1994; Koerner \& Sargent 1995).
The merit of using CO lines is that the molecule is in general much more abundant
than any other molecules with heavy elements in interstellar conditions.
However, the number of disks observed via molecular lines is much smaller than
that via dust continuum studies.  One of the difficulties in searching
for gaseous disks via CO lines is contamination of the lines with ambient
cloud gas. CO is abundant not only in the disk, but also in ambient
clouds, and the critical densities of the excitation of its $J=2-1$ and
$J=1-0$ lines are comparable to the cloud gas density.
Hence it is useful if we can find molecular lines which 
preferentially trace the gas in disks. Such lines have to satisfy the
following two conditions: (a) their critical densities should be
higher than a typical density of molecular clouds ($n_{\rm H}\sim 10^4$
cm$^{-3}$) and (b) the molecule should be reasonably abundant within the
disk. In fact, there are many molecular rotational lines, especially
the high$-J$ lines, the critical density of which is higher than the cloud
density but lower than or comparable to the disk density.
The latter condition (b) can be checked based on our model.

In our calculation of molecular column densities in disks in Paper I, however,
we assumed the disks to be directly exposed to the interstellar radiation
field. In order to see if there is any molecular line which can be used
to trace embedded disks, we have to reconsider the problem including
the UV attenuation via ambient gas. One possible tracer is CN,
the rotational ($N=2-1$) transition of which is one
of the strongest lines detected in the disks around DM Tau, GG Tau, and
LkCa15 (Dutrey et al. 1997; Qi 2000).
The clear detection of the line indicates that CN is relatively abundant
in these disks, and the high critical densities ($n_{\rm H}=3\;10^5$
cm$^{-3}$ for $N=1-0$ and $ 10^6$ cm$^{-3}$ for $N=2-1$)
of the lines seem ideal in order to avoid
contamination with ambient clouds at lower density. But theoretical
studies show that CN exists mostly in the surface regions of disks, in which
interstellar UV plays a dominant role in the chemistry (Paper I), and thus
it is not clear if it is also abundant in disks shielded from
interstellar radiation.
In this section, we discuss the column densities of CN, together with
other species, in  embedded disks.

Fig.  \ref{fig:atten_noX} and \ref{fig:atten_X} show calculated
radial column density  distributions of assorted species  in
the lower mass disk model at $t=9.5\; 10^5$ yr.  Fig.
\ref{fig:atten_noX} contains model results in the absence of any X-ray
emission from the central star, while Fig.  \ref{fig:atten_X} shows
results when X-ray processes are included.  The solid lines, dotted
lines, and dashed lines are for $A_{\rm v}$ = 0, 1, and 2 mag,
respectively.  It is easy to see that the distributions can
depend on both the X-ray irradiation and the degree of extinction.
In the model without X-rays, the column densities of
radicals such as CN and C$_2$H are extremely sensitive to the
existence of ambient gas above and below the disk since this gas
interferes with the photodissociation that produces the radicals.  The
effect is especially drastic at inner ($R\lesssim 300$ AU) radii,
where the gas density is higher and thus abundances of radical species
are very low without photoprocesses.  It should be noted, however,
that the column density of CN in the case with $A_{\rm v}=2$ mag is
still reasonably large at $R\gtrsim 300$ AU. We additionally
calculated molecular abundances at $R=500$ AU for $A_{\rm v}=4$ mag,
in which the CN column density is $3.7\; 10^{12}$ cm$^{-2}$.  These
results suggest that CN can be a disk tracer at least in the outer
regions ($R\gtrsim 300$ AU) for T Tauri stars without high X-ray
luminosity.

In the model with X-rays from the central star, the column densities of
radicals such as CN, CH, and C$_2$H are high
even in the inner regions.
The high ionization rate and induced photolysis by X-rays significantly
enhance the radical species in regions with small radius and large height.
We should note that the column densities of those radicals
obtained in this model are
upper limits because the X-ray luminosity we assumed --
$L_{\rm x}= 10^{31}$ erg
s$^{-1}$ -- is close to the upper limit of the temporally varying
X-ray intensity from T Tauri stars.  But sufficient column densities
of the radicals are expected in the inner regions ($\lesssim 200$ AU)
even if the average X-ray luminosity is lower by an
order of magnitude, since their radical column densities roughly depend
linearly on the X-ray luminosity there.
Therefore CN can trace even the inner regions ($R \lesssim 200$ AU) of
embedded disks when there is some X-ray luminosity.

As for other species, it is interesting to note that the column
densities of HCO$^+$ and NH$_3$ tend to be higher in disks with
non-zero $A_{\rm v}$, regardless of the X-ray luminosity, which
suggests that their high frequency transitions with high critical
densities can also be tracers of embedded disks. The centrally peaked
distribution of HCO$^+$ makes it the better tracer because the line
can be more sensitive to a disk with small radius.

Although the calculations reported in this section deal with the
effect of embedding disks in clouds, a close observation of Figs.
\ref{fig:atten_noX} and \ref{fig:atten_X} shows that some qualitative
trends are actually independent of the degree of extinction.  For
example, species such as CN, C$_2$H, and H$_2$CO show a central peak
in the case with X-ray irradiation and a central hole in the case of
no such irradiation regardless of whether or not the disk is embedded.
This behavior is not universal; the HCN distribution shows a
similar dependence on the X-ray luminosity only when the disks
are embedded.  Specifically, in the model without X-ray irradiation,
HCN (in the case of $A_{\rm v}\ge 1$ mag) shows a central hole of
radius $\sim 200$ AU, while in all cases with X-ray irradiation it is
centrally peaked. Thus, if X-ray fluxes are different from one
disk to another, one can expect the radial distributions of
individual molecules to be different. Interestingly,
interferometric observations show that CN is centrally peaked in the
disk around DM Tau (Dutrey 2000, personal communication) but exhibits
a central hole in the disk around LkCa15 (Qi 2000).


\section{Summary}
We have presented two major extensions from our previous models
reported in Paper I.

Firstly, we have investigated the two-dimensional distribution of
deuterated species in protoplanetary disks. The molecular D/H ratios
in gaseous disks evolve in a similar way as in molecular clouds.
In the vertical direction the D/H ratios are higher at lower height,
where molecules are more heavily depleted onto grain surfaces.

The D/H ratios
of HCO$^+$, NH$_3$, and H$_2$O decrease significantly towards smaller
radii at $R\lesssim 300$ AU. These molecules are deuterated through
H$_2$D$^+$, the abundance of which is rather sensitive to the
temperature, which increases at small radii.  The D/H ratios of CH$_4$
and H$_2$CO do not decrease inwards because these species are
deuterated through CH$_2$D$^+$, which is formed from CH$_{3}^{+}$
through a sufficiently exothermic reaction that it does not possess
much of a temperature dependence in the range ($T\lesssim 40$ K) we are
concerned with in this paper.
The DCN/HCN ratio is higher at
larger radii because DCN is formed primarily from H$_2$CN via the
reaction with D atoms, which are more abundant at larger radii, where
the density is lower.

The D/H ratios depend not only on the temperature, but also on the disk
mass and the X-ray flux.
In the disk with larger mass, the column-density ratios of deuterated
species to normal species are generally higher because of 
heavier depletion
of molecules onto grains.  With a higher X-ray flux, the D/H ratios
of  HCO$^+$, NH$_3$ and H$_2$O are smaller at large radial distances
  because the enhancement of the H$_2$D$^+$
to H$_3^+$ ratio is reduced by more copious electrons.

Our results for the ratio of the column densities of DCN and HCN
are in reasonable agreement with the observation of the disk around
LkCa15 (Qi 2000) whether we utilize the Kyoto or the low mass model, but
the absolute column densities we calculate are too low for this
rather massive disk.

Secondly, we have investigated molecular column densities in disks
embedded in ambient clouds using models without and with X-rays from
the central T Tauri star.

In the models without X-rays, the column
densities of radicals such as CN are sensitive to the attenuation of
interstellar UV by ambient gas, especially in the inner radius of the disk.
In the disk with direct UV irradiation, the CN column density
is $3\; 10^{12}-3\;10^{13}$ cm$^{-2}$ in the region $50\le R \le 700$
AU, while
in the disk embedded in ambient gas, the column density of CN 
is $10^{12} - 10^{13}$
cm$^{-2}$ in the outer region of the disk ($R\gtrsim 300$ AU), but
decreases for radii
  $\lesssim 300$ AU by orders of magnitude.
On the other hand, in the models with X-rays, ionization and induced
photolysis by X-rays enhance the abundance of CN and other
radicals so that the column density of CN
is as high as $\gtrsim 10^{13}$ cm$^{-2}$ at $50\le R \le 700$ AU.
Since the X-ray flux is higher in inner regions, radial distributions
of radicals such as CN are centrally peaked.  Indeed, the distributions
for non-radicals tend to show a similar effect.

The molecular ion HCO$^+$ is centrally peaked regardless of the X-ray
luminosity, and is more abundant in the case with UV shielding via
ambient clouds. Inclusion of X-rays further enhances the column density of
HCO$^+$.

Our calculated results, combined with the fact that CN and HCO$^+$ are clearly
detected in the non-embedded disks around DM Tau, GG Tau, and LkCa15,
suggest that CN and HCO$^+$ can be gaseous disk tracers towards embedded
objects especially with high X-ray luminosity.
Our results also suggest that the radial distribution of the
radicals CN and C$_2$H and some non-radical species (HCN and H$_2$CO)
may vary among disks depending on the X-ray luminosity of the central
star.

\acknowledgements
The authors are grateful to A. Dutrey, G. Blake, C. Qi, and M. Saito
for stimulating discussions on the line observation of protoplanetary disks.
Y. A. is grateful for financial support from the Japan Society for Promotion
of Science. The Astrochemistry Program at The Ohio State
University is supported by The National Science Foundation.
Numerical calculations were partly carried out at the Astronomical Data
Analysis Center of the National Astronomical Observatory of Japan, and
on the Cray T90 at the Ohio Supercomputer Center.



\begin{thebibliography}{}

\bibitem[]{} Adams, F. C., \&  Lin, D. N. C. 1993, in  Protostars and Planets 
III, ed. E. H. Levy, \& J. I. Lunine (Tucson: 
 Univ. of Arizona Press), 721
\bibitem[]{} Aikawa, Y.,  \& Herbst, E. 1999a, A\&A,  351, 233 (Paper I)
\bibitem[]{} Aikawa, Y.,  \& Herbst, E. 1999b, ApJ, 526, 314
\bibitem[]{} Aikawa, Y., Umebayashi, T., Nakano, T., \&  Miyama, S. M.,
1997, ApJ, 486, L51
\bibitem[]{} Aikawa, Y., Umebayashi, T., Nakano, T.,  \& Miyama, S. M. 1999,
ApJ, 519, 705
\bibitem[]{} Anders, E., \& Grevesse, N. 1989, Geochimica et Cosmochimica
Acta, 53, 197
\bibitem[]{} Beckwith, S. V. W., Sargent, A. I., Chini, R. S., \& 
G\"usten, R. 1990, AJ, 99, 924
\bibitem[]{} Brown, P. D., \& Millar, T. J. 1989, MNRAS, 237, 661
\bibitem[]{} Cameron,  A. G. W. 1973, Space Science Reviews, 15, 121
\bibitem[]{} Caselli,  P., Walmsley,  C. M., Terzieva,  R., \& Herbst, E. 1998,
ApJ, 499, 234
\bibitem[]{} Dutrey, A., Guilloteau, S.,  \& Gu$\acute{\rm e}$lin, M.
1997, A\&A, 317, L55
\bibitem[]{} Dutrey, A., Guilloteau, S., \& Simon, M. 1994, A\&A,
286, 149
\bibitem[]{} Glassgold, A. E., Najita, J.,  \& Igea J. 1997,
ApJ, 480, 344
\bibitem[]{} Gredel, R., Lepp, S., \& Dalgarno, A. 1987, ApJ, 323, L137
\bibitem[]{} Gredel, R., Lepp, S., Dalgarno, A., \& Herbst, E. 1989, ApJ,
 347, 289
\bibitem[]{} Gu$\acute{\rm e}$lin, M., Langer, W. D., \& Wilson R. W. 1982, A\&A,
107, 10
\bibitem[]{} Guilloteau,  S.,  \& Dutrey, A. 1998, A\&A, 339, 467
\bibitem[]{} Handa, T., Miyama, S. M., Yamashita, T., et al. 1995, ApJ, 449, 894
\bibitem[]{} Hayashi, C. 1981, Prog. Theor. Phys. Suppl., 70, 35
\bibitem[]{} Henchman, M. J., Paulson, J. F., Smith, D., Adams N. G., 
\& Lindinger, W.
1988, in Rate Coefficients in Astrochemistry, ed. T. J. Millar, \& D.
A. Williams (Dordrecht: Kluwer), 201 
\bibitem[]{} Herbig, G. H., \& Goodrich, R. W. 1986, ApJ, 309, 294
\bibitem[]{} Imhoff, C. L., \& Appenzeller, I. 1987, in Scientific 
Accomplishments of the I. U. E., ed. Y. Kondo (Dordrecht: Reidel),  295
\bibitem[]{} Kawabe, R., Ishiguro, M., Omodaka, T., Kitamura, Y., \&
Miyama, S. M. 1993, ApJ, 404, L63
\bibitem[]{} Koerner, D. W., \& Sargent, A. I. 1995, AJ, 109, 2138
\bibitem[]{} Koerner, D. W., Sargent, A. I.,  \& Beckwith, S. V. W. 1993,
Icarus, 106, 2
\bibitem[]{} Lee, H.-H., Herbst, E., Pineau des For\^{e}ts, G., Roueff, E.,
  \&  Le Bourlot, J. 1996, A\&A, 311, 690
\bibitem[]{} Lee H.-H., Roueff, E., Pineau des For\^{e}ts G., et al.
1998, A\&A, 334, 1047
\bibitem[]{} Lennon, M. A., Bell, K. L., Gilbody, H. B., et al.
1988, J. Phys. Chem. Ref. Data, 17, 1285
\bibitem[]{} Maloney,  P., Hollenbach, D. J.,  \& Tielens, A. G. G. M. 1996,
ApJ, 466, 561
\bibitem[]{} Millar, T. J., Bennett, A., \& Herbst, E. 1989, ApJ, 340, 906
\bibitem[]{} Millar, T. J., Farquhar, P. R. A., \& Willacy, K. 1997, 
A\&AS, 121, 139
\bibitem[]{} Montmerle, T., Feigelson, E. D., Bouvier, J., \&
Andr$\acute {\rm e}$, P. 1993, in Protostars and Planets III, ed. E. 
H. Levy, \& J. I. Lunine (Tucson: Univ. of Arizona Press), 689
\bibitem[]{} Nesbitt, F. L., Marston, G., \& Stief, L. J. 1990,
J. Phys. Chem., 94, 4946
\bibitem[]{} Osamura, Y., Fukuzawa, K., Terzieva, R., \& Herbst, E. 1999,
ApJ, 519, 697
\bibitem[]{} Qi, C. 2000, PhD thesis, California Institute of
Technology
\bibitem[]{} Qi, C., Blake, G. A., \& Sargent, A. I. 1999, in Science with
the Atacama Large Millimeter Array (ALMA), a meeting held October 6-8, 1999 at
Carnegie Institution of Washington
\bibitem[]{} Ruffle, D. P., Hartquist, T. W., Taylor, S. D., \&
   Williams, D. A. 1997, MNRAS, 291, 235
\bibitem[]{} Saito, M., Kawabe, R., Ishiguru, M., et al. 1995, ApJ, 453, 384
\bibitem[]{} Scott, G. B. I., Fairley, D. A., Freeman, C. G., McEwan, M. J., 
\& Anicich, V. G. 1998, J. Chem. Phys., 109, 9010
\bibitem[]{} Terzieva, R.,  \& Herbst, E. 1998, ApJ, 501, 207
\bibitem[]{} van der Tak, F. F. S., \& van Dishoeck, E. F. 2000, A\&A, 
358, L79
\bibitem[]{} van Dishoeck, E. F., \&  Black, J. H. 1988, ApJ, 334, 771
\bibitem[]{} Verner, D. A., Yakovlev, D. G., Band, I. M., \& Trzhaskovskaya, M. B.
1993, At. Data Nucl. Data Tables, 55, 233
\bibitem[]{} Willacy, K., \& Langer, W. D. 2000, ApJ, 544, 903

\end{thebibliography}
\end{document}